\documentclass[12pt]{article}

\input{psfig}

\newcommand{\bibl}[5]
	{#1, {\it #2} {\bf #3} #5 (#4)}
\newcommand{\eabe}{\begin{eqnarray}}
\newcommand{\eaen}{\end{eqnarray}}
\newcommand{\eqbe}{\begin{equation}}
\newcommand{\eqen}{\end{equation}}

\begin{document}

\begin{titlepage}
\begin{flushright}
  LU TP 96-8 \\
  February 1996
\end{flushright}
\vspace{25mm}
\Large
\begin{center}
  {\bf Colour: A Computer Program for QCD Colour Factor Calculations} \\
  \normalsize
  \vspace{12mm}
  Jari H\"akkinen, Hamid Kharraziha\footnote{jari@thep.lu.se, hamid@thep.lu.se} \\
  Department of Theoretical Physics, Lund University, \\
  S\"olvegatan 14A, SE-223 62 Lund, Sweden \\
  \vspace{10ex}
  Submitted to Computer Physics Communications
\end{center}
\vspace{6ex}
\normalsize
{\bf Abstract:} \\
A computer program for evaluating colour factors of QCD Feynman diagrams is presented, and illustrative examples on how to use the program to calculate non trivial colour factors are given. The program and the discussion in this paper is based on a diagrammatic approach to colour factors.
\end{titlepage}

\normalsize
\section*{Program Summary}
{\em Title of program:} colour

{\em Program obtainable from:} http://www.thep.lu.se/tf2/hep \\
\mbox{} \hspace{9.6em} {\em or} ftp:thep.lu.se/pub/LundPrograms/Misc/colour.tar.Z

{\em Computer for which program is designed:} Computers with an ANSI C compiler.

{\em Program language used:} C

{\em Peripherals used:} terminal and mass storage for input, terminal for output

{\em Number of lines in combined program:} $\approx$1000

{\em Keywords:} Feynman diagrams, QCD colour factors

{\em Nature of the physical problem:} In many QCD Feynman diagrams there is a need to calculate the colour factor. This program calculates colour factors fast and accurately.

{\em Method of solution:} QCD Feynman diagrams, with no four-gluon vertices, factorize into a colour (group) factor and a kinematical factor. The colour part of any closed QCD Feynman diagram can be transformed into a sum of diagrams including only closed quark (colour) lines. In each such term, the number of quark (colour) loops is counted, giving factors of 3 (or $N_c$). The sum of these terms is then the desired colour factor. \\
To perform the translation, we need to have rules to interpret gluons and vertices into quark lines. There is in principle only need for three equalities to be able to calculate any closed QCD diagram (not containing four gluon vertices). To achieve good computing performance further equalities have been introduced. The most essential ones are listed in Appendix~\ref{a:rules}.

{\em Restriction of complexity of the problem:} Four gluon vertices cannot be included directly in diagrams to be calculated. The number of gluons in diagrams to be calculated are limited to 200, but can easily be changed prior to compiling.

{\em Typical running time:} For typical diagrams, fractions of a second. Complicated diagrams - a few seconds to hours.

\newpage
\section{Introduction} \vspace{-2ex}
There are several programs available to perform momentum Feynman diagram calculations, either symbolically or numerically~\cite{hs74}. Some of these perform QCD calculations, but usually the QCD calculating programs do not calculate the colour factors for a given diagram. The users have to calculate these factors themselves. This can be done in different ways, the most appealing method is presented in~\cite{russia,pc84} where only diagrammatic manipulations are needed to calculate colour factors of diagrams.

In this paper we present a computer program for calculating colour factors of QCD diagrams using the rules and equalities derived in~\cite{russia}. An introduction is given on how to use the program and which results can be obtained. Using a set of projection operators non trivial results can be achieved, such as how the interaction probability between two gluons depend on their common colour state. Only closed Feynman diagrams (vacuum QCD diagrams, ie. diagrams with no external legs) can be calculated, but this is not a real problem since the restriction can be circumvented as will be shown. We also present a short introduction to the group theory part of the calculations.

\section{Theory in brief} \vspace{-2ex}
The program presented here is based on the results in~\cite{russia} and~\cite{pc84}. These results has been further developed and explained to us by Yuri Dokshitzer and Bo S\"oderberg~\cite{yd92}. In this section and the appendices we give a brief presentation of the main results.

The existence of colour factors in QCD originates from the invariance of
the Lagrangian under colour transformations. For a certain interaction, with
incoming and outgoing particles, intermediate propagators and different
kinds of vertices, one has to sum over all possible colours, if they
are not explicitly specified. Since the kinematical factor in the interaction
probability is equal for all these diagrams (excluding diagrams containing four-gluon vertices), we therefore have a multiplicative factor to the probability, which we get by counting the colours. 

Two preliminary remarks. When we discuss colours and anti-colours we often use quarks and anti-quarks, respectively, to describe the colour flow. In most expressions we keep $N_c$ as a free variable in order to be general. Whenever the number of colours is specified, $N_c=3$ has been used. 

Every $3^m \otimes \bar{3}^n$ space can be divided into a set of irreducible subspaces. In the simplest example of a quark and an anti-quark system, the irreducible representation is the singlet and the octet state, ie. $3 \otimes \bar{3} = 1 \oplus 8$, where the octet state corresponds to the gluons. The equation
\eqbe
\label{e:3x3}
\psfig{figure=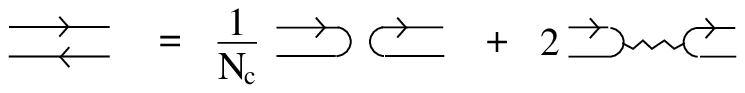}
\eqen
states this fact in equation form. (The factor of 2 in the second term is due to the Gell-Mann normalization of gluons, see discussion in Appendix~\ref{a:operators}. We have adopted Gell-Mann normalization in all calculations in this paper.) In the LHS of Eq~(\ref{e:3x3}) a colour and an anti-colour state, $3 \otimes \bar{3}$, are divided into irreducible singlet and octet states on the RHS. The degeneracy (multiplicity) of a state is given by taking the trace of the representation of the state. Calculating the trace in this diagrammatic representation corresponds to connecting the incoming legs and the outgoing colour flow. For the LHS we find
$$
\psfig{figure=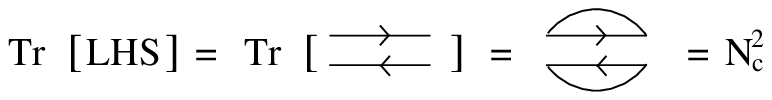}
$$
as expected, and for the {RHS}
$$
\psfig{figure=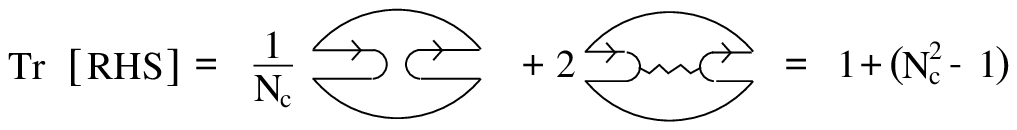}
$$

Another observation is that the two diagrams, $P_1$ and $P_2$, in the RHS of Eq~(\ref{e:3x3}) are disjoint projection operators since the terms satisfy $P_i^2 = P_i$ and $P_i P_j =0, P_j P_i = 0$. These can be used to split any quark--anti-quark diagram into an irreducible representation
$$
\psfig{figure=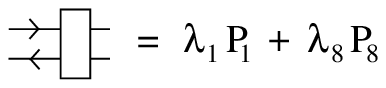}
$$
The projection operators can be used to ask specific questions about the colour states particles are in, eg. what is the probability to find two free gluons in an anti-symmetric octet state. This is easily solved by calculating colour states. The two gluons are in a $8 \otimes 8$ state with degeneracy 64, and the irreducible representation of the $8 \otimes 8$ space contains an anti-symmetric octet (with degeneracy 8), ie. the answer is 8/64 since the probability is equal for all states. In the next section we will give a few other examples on these issues and describe how to perform the calculations using the program. 

Given a problem concerning colour states, the problem of finding solutions boils down to finding the projection operators for reduced representations. Once this is done all interesting numbers can be calculated using the program. In Appendix~\ref{a:operators} we have listed the (irreducible representation) projection operators for a few states.

The triple gluon coupling is anti-symmetric in colour. We have generalized the program to include also a symmetric coupling between three colour octets (``gluons'').

\section{Using the program} \vspace{-2ex}
\label{s:useprog}
A straight forward use of the program is to calculate QCD vacuum diagrams. This might not be too exciting. Using projection operators more interesting results can be obtained. A few illustrative examples will be given below.

The program interacts with the user through input files. The diagram to be calculated is defined in one input file which is read by the program, the diagram is calculated and the result is written on the screen. No interaction with the user is needed after the program has been started. 

The command line syntax to start the program is: \vspace{-2ex}
\begin{center}
colour [-df file $|$ -h $|$ -help] \vspace{-3ex}
\end{center}
where \vspace{-3ex}
\begin{description}
  \item[\hspace{1em} -df] Changes the input file name. The default name is {\em colour.in}.  \vspace{-2ex}
  \item[\hspace{1em} -h] or {\bf -help} Prints a small description of the program and a list of valid options.
\end{description}
There are few rules for defining a qcd diagram in the input file. \vspace{-3ex}
\begin{itemize}
  \item Every gluon line is assigned a unique number. \vspace{-2ex}
  \item Triple anti-symmetric gluon vertices are represented by a letter $f$ followed by the three incoming gluon numbers. \vspace{-2ex}
  \item Symmetric triple vertices\footnote{As mentioned above these are not physical QCD gluon vertices, but are included for generality. The vertex is also used in projection operators.} are represented in the same way as the anti-symmetric vertices but with the letter $d$. \vspace{-2ex}
  \item Quark loops are represented by a letter $t$, followed by all gluons connected to the quark loop. \vspace{-2ex}
  \item Four-gluon vertices cannot be directly included in the program since the colour and the kinematical parts of the cross section do not factorize. This can, however, be taken care of by separating the four-gluon vertex in factorizable terms, as discussed further in Appendix~\ref{a:rules}. \vspace{-2ex}
  \item Ghosts can also be included since these couple to gluons with the same colour factor as gluons themselves. Thus ghost vertices can be represented with a letter $f$. \vspace{-2ex}
  \item For the quark loops and the anti-symmetric gluon vertex the order of the gluons are important since $f_{abc} = -f_{bac}$. So when translating a diagram into a file all loops and $f$ vertices must be written down in the same order (clockwise or anti-clockwise). \vspace{-2ex}
  \item Composite objects, d, f, and t, must be separated with a newline character, while gluon numbers are separated with a space. All other separation characters are invalid and will stop the execution of the program.
\end{itemize}

For example, the diagram
$$
\psfig{figure=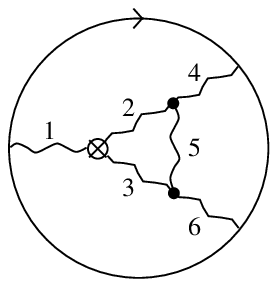}
$$
would transform to a file like \vspace{-3ex}
\begin{tabbing}
\hspace{2em} \= d \= 1 2 3 \\
\> f \> 2 4 5 \\
\> f \> 3 5 6 \\
\> t \> 1 4 6
\end{tabbing} \vspace{-3ex}
The numbers assigned to every gluon are arbitrary and occur twice in the file since every gluon is connected to two vertices in the diagram. If you run this file, the answer will be: $\frac{1}{2^3}(N_c^4-5N_c^2+4)$ and is equal to 5 using $N_c=3$.

\subsection{Example 1: Colour factor for gluon emission from quarks} \vspace{-2ex}
\label{ss:ex1}
Let us start with the problem of calculating the colour factor, often denoted 
$\rm{C_F}$,
for emission of a gluon from a quark line. This can be easily done
by hand, but it is a nice example to demonstrate how the program
can be used. 

\begin{figure}[t,b]
  \hbox{\vbox{
    \begin{center}
    \mbox{\psfig{figure=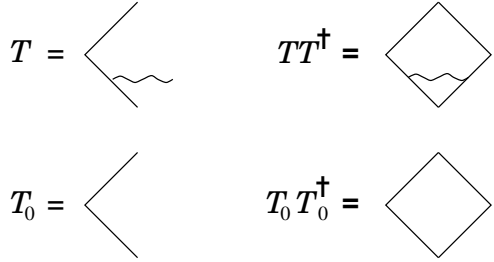}}
    \end{center}
  }}
  \caption{\em Diagrams used to calculate $C_F$.}
  \label{f:ex1}
\end{figure}
We will study a transition matrix T, defined in Fig~\ref{f:ex1}. A
 colour singlet comes in, a quark--anti-quark pair is created and
a gluon is emitted from one of the quark legs.
We have chosen to use a singlet state since the program can only handle closed diagrams\footnote{There are obviously other diagrams which would enable us to calculate $C_F$, but these involve obscuring technicalities.}. 
By definition, $\rm{C_F}$ is a 
multiplicative factor to the transition probability, due to the 
emission of a gluon from the quark leg. For the transition probability,
 we have to square the matrix by multiplying it
with its conjugate (reversing the colour flow and using the mirror image).
The result then must be normalized by dividing by
the squared amplitude of just creating the quark--anti-quark pair.

Consequently, to obtain $C_F$
we have to run the program twice, one time with the diagram $TT^\dagger$
and once with the diagram $T_0T_0^\dagger$.

In order to use the program we have to translate the diagrams into an input file. This can be done as \vspace{-3ex}
\begin{center}
\begin{tabular}{c|l|c}
Diagram		 & Input file	& Result \\
\hline
$TT^\dagger$	 & t 1 1 	& $\frac{1}{2} \left[N^2_c - 1 \right]$ \\
\hline
$T_0T_0^\dagger$ & t		& $N_c$
\end{tabular}
\end{center}
\vspace{-2ex} Thus, we get that 
\[C_F=\frac{N^2_c - 1}{2 N_c} = \frac{4}{3}. \]
 
\subsection{Example 2: Partitioning into irreducible representations} 
\begin{figure}[t,b]
  \hbox{\vbox{
    \begin{center}
    \mbox{\psfig{figure=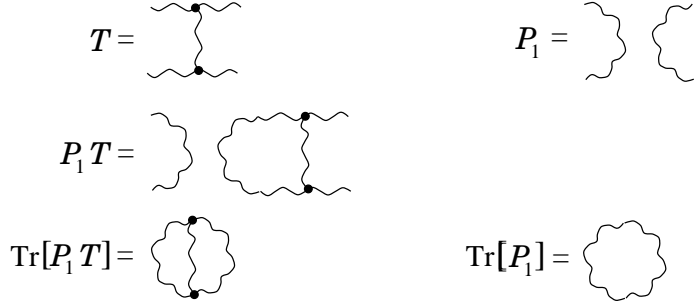}}
    \end{center}
  }}
  \caption{\em Diagrams used to calculate the singlet eigenvalue, $\lambda_1$, for the state $T$.}
  \label{f:ex2}
\end{figure}

\vspace{-2ex}
\label{ss:ex2}
Any diagram in the $8 \otimes 8$ space can be written as a sum of the projection
operators described in Appendix~\ref{a:operators} (and crossing operators which are built from operators of the same multiplicity). Consider eg the gluon-gluon scattering matrix T in Fig~\ref{f:ex2}. This matrix can be written as a sum of
projection operators solely, that is without any crossing operators,
$$
T=\sum \lambda_i P_i.
$$
The eigenvalues can easily be found. To be concrete we will determine $\lambda_1$ corresponding to the singlet state, and the other eigenvalues are derived in the same way. To
calculate this, we apply the operator to the matrix T, thus projecting
out the singlet part of the diagram. Then we take the trace of $PT$ and
divide the result with the trace of just the operator,
$$
TP_1=\lambda_1P_1,\ \ \ \ \ \lambda_1=\frac{\mbox{Tr}(P_1T)}{\mbox{Tr}(P_1)}.
$$

The translation into a file looks like \vspace{-2ex}
\begin{center}
\begin{tabular}{c|l|l}
Diagram		 	& Input file	& Result \\
\hline
$\mbox{Tr}(P_1T)$	& f 1 2 3 	& \\
			& f 1 3 2	& $N_c \left[N^2_c - 1 \right]$ \\
\hline
$\mbox{Tr}(P_1)$	& See Sec~\ref{ss:limits}& $N^2_c - 1$ \\
\end{tabular}
\end{center}
\vspace{-2ex} Running the program we find that $\lambda_1=N_c$.\footnote{Note that the eigenvalues depend on the normalization of the gluon, see Appendix~\ref{a:operators}.}
 
\subsection{Example 3: Probability of an octet state} 
\begin{figure}[t,b]
  \hbox{\vbox{
    \begin{center}
    \mbox{\psfig{figure=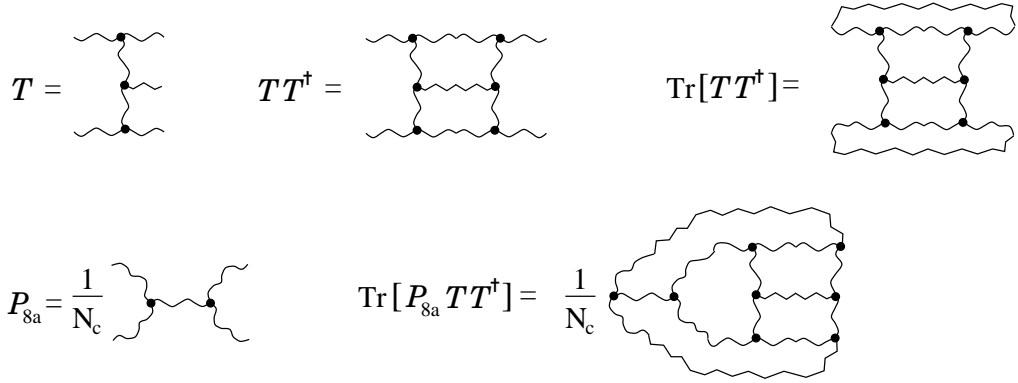}}
    \end{center}
  }}
  \caption{\em Diagrams used in Example 3.3.}
  \label{f:ex3}
\end{figure}

\vspace{-2ex}
\label{ss:ex3}
The question here is: If we have a transition matrix of some kind containing two gluons, what is the probability for them to be in any of the states described in Fig~\ref{f:8x8operators} in Appendix~\ref{a:operators}. Take for example the transition matrix T in Fig~\ref{f:ex3}, what is the probability for the two incoming gluons to be in the anti-symmetric octet state $P_{8a}$?

To solve this problem, the first step is to square the matrix, thus getting the symmetric matrix $TT^\dagger$ which correspond to the probability. Then the antisymmetric octet part of the amplitude is projected out, by applying the operator. The probability is given by the multiplicity of the projected part, divided by the total multiplicity of the amplitude,
$$
	Prob(P_{8a}|T)=\frac{Tr(P_{8a}TT^\dagger)}
	                    {Tr(TT^\dagger)}.
$$
For this we have to run the program twice, once with the diagram $Tr(TT^\dagger)$ and once with the diagram $Tr(P_{8a}TT^\dagger)$.

The translation into a file looks like \vspace{-2ex}
\begin{center}
\begin{tabular}{c|l|l}
Diagram		 	& Input file	& Result \\
\hline
			& f 1 4 2	& \\
			& f 4 7 5	& \\
			& f 7 9 8	& \\
			& f 6 8 9	& \\
			& f 3 5 6	& \vspace{-12.5ex}\\
\psfig{figure=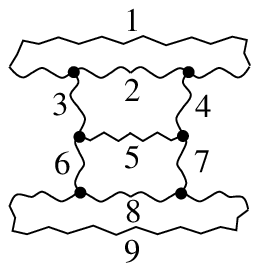} & f 1 2 3 & $N^3_c \left[N^2_c - 1 \right]$ \\
\hline
$N_c \, \mbox{Tr}(P_{8a}TT^\dagger$)	& f 1 4 2	& \\
			& f 4 7 5	& \\
			& f 7 9 8	& \\
			& f 6 8 12	& \\
			& f 3 5 6	& \\
			& f 11 2 3	& \\
			& f 1 10 9	& \\
			& f 10 11 12	& $\frac{1}{8}\left[N^6_c + 11 N^4_c - 12 N^2_c \right]$ \\
\end{tabular}
\end{center}
\vspace{-2ex} Putting it all together gives a probability of 7/24 (with $N_c =3$) to find the two incoming gluons in an octet state.

\subsection{Limitations}
\label{ss:limits}
There are some limits in the program. First, the physics limitations. Four-gluon vertices are not included and therefore diagrams containing these vertices cannot be calculated directly. The four-gluon vertex must be replaced with three other diagrams, which have to be calculated separately (see Appendix~\ref{a:rules}). There is also one specific diagram that the program cannot evaluate, a gluon ring, but the colour factor for this diagram is simply the number of different gluons, $N^2_c - 1$. This is purely technical and the diagram could in principle be implemented into the program as a special case, but the gain is not worth the effort. \\
Secondly, local implementation limits. There is a limit on the number of gluons which can be included in the program. This is set to 200 but can easily be changed by the user before compiling. There is also a limit to which power in $N_c$ calculations can be done, the default range is $N^{-100}_c$ -- $N^{99}_c$, this can also easily be changed prior to compiling the program. A warning message will appear if over- or underflow occurs.

\section{Acknowledgments} \vspace{-2ex}
We would like to thank Yuri Dokshitzer and Bo S\"oderberg for introducing us into the world of colour factors.

\appendix
\newpage
\section{Colour Rules}
\label{a:rules}
\begin{figure}[h]
  \hbox{\vbox{
    \begin{center}
    \mbox{\psfig{figure=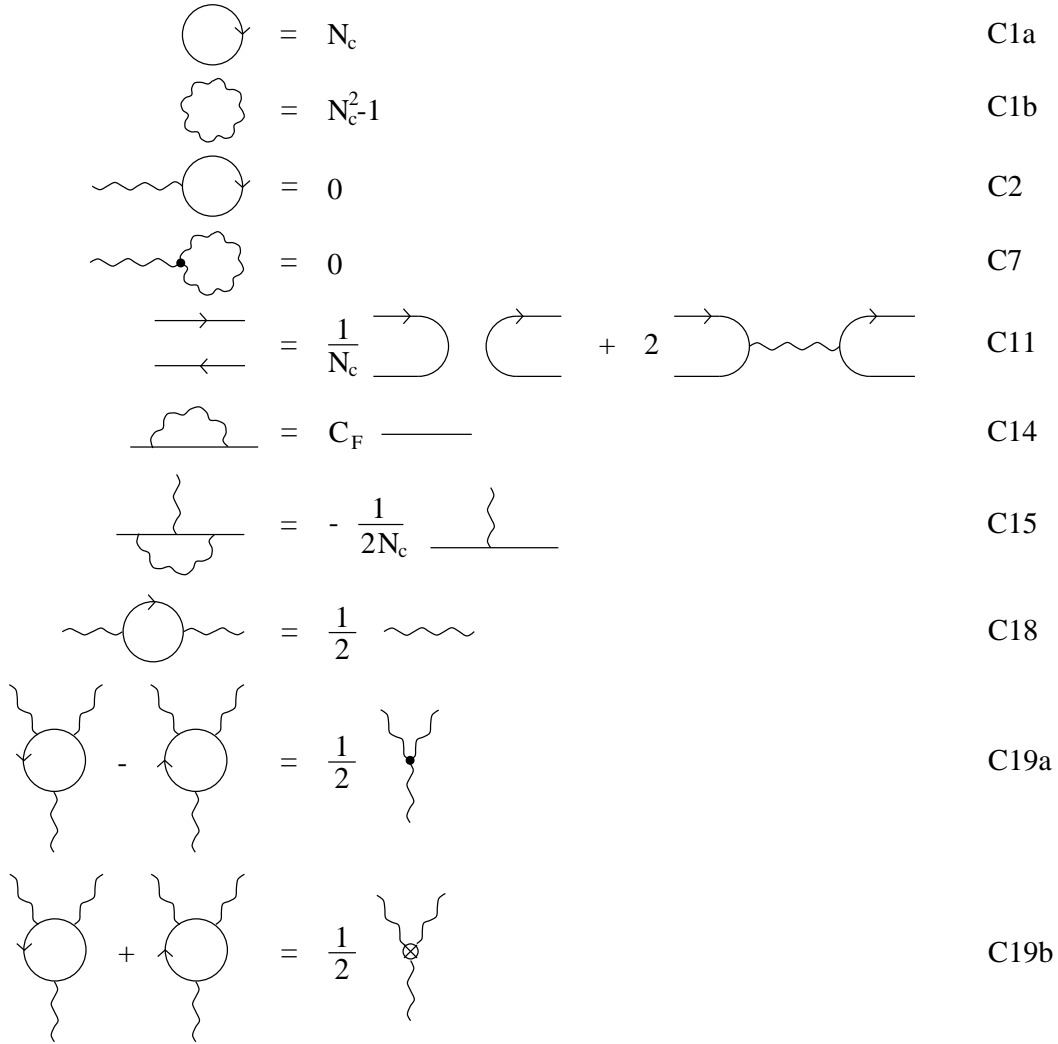}}
    \end{center}
  }}
  \caption{\em Rules for diagram manipulations.}
  \label{f:rules}
\end{figure}

The set of rules in Fig~\ref{f:rules}, taken from~\cite{russia}, are implemented in the program with the exception of the gluon loop C1b. The equation numbers are the ones to be found in~\cite{russia}, and as indicated we do not list all of them. We note that in principle it is enough to use equations C11 and C19 to translate any diagram without four-gluon vertices to diagrams containing only quark loops. The other equalities are implemented in order to increase the computing performance.

\begin{figure}[h]
  \hbox{\vbox{
    \begin{center}
    \mbox{\psfig{figure=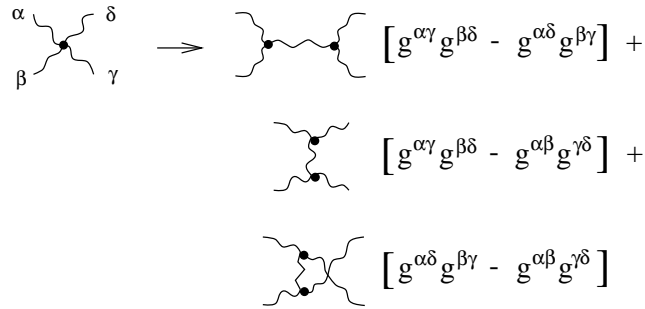}}
    \end{center}
  }}
  \caption{\em Symbolic representation of the colour factors in a four-gluon vertex.}
  \label{f:4-gluon}
\end{figure}

The four-gluon vertex has not been included in the program since QCD diagrams containing four-gluon vertices do not factorize into a colour (group theoretical) part and a kinematical part. The four-gluon vertex,
$$
\psfig{figure=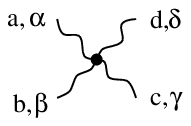}
$$
has the form 
$$
f_{abn} f_{ncd} [g^{\alpha \gamma} g^{\beta \delta} - g^{\alpha \delta} g^{\beta \gamma}] + f_{and} f_{bcn} [g^{\alpha \gamma} g^{\beta \delta} - g^{\alpha \beta} g^{\gamma \delta}] + f_{anc} f_{bdn} [g^{\alpha \delta} g^{\beta \gamma} - g^{\alpha \beta} g^{\gamma \delta}].
$$
Thus a diagram with such a vertex can be split into three pieces, each of which has a factorized form (this separation is not unique). Each piece has a colour factor corresponding to a diagram where the four-gluon vertex is replaced by two three-gluon vertices, as symbolically shown in Fig~\ref{f:4-gluon}.

\section{Projection Operators}
\label{a:operators}
\begin{figure}[b]
  \hbox{ \hspace{-2em}
    \mbox{\psfig{figure=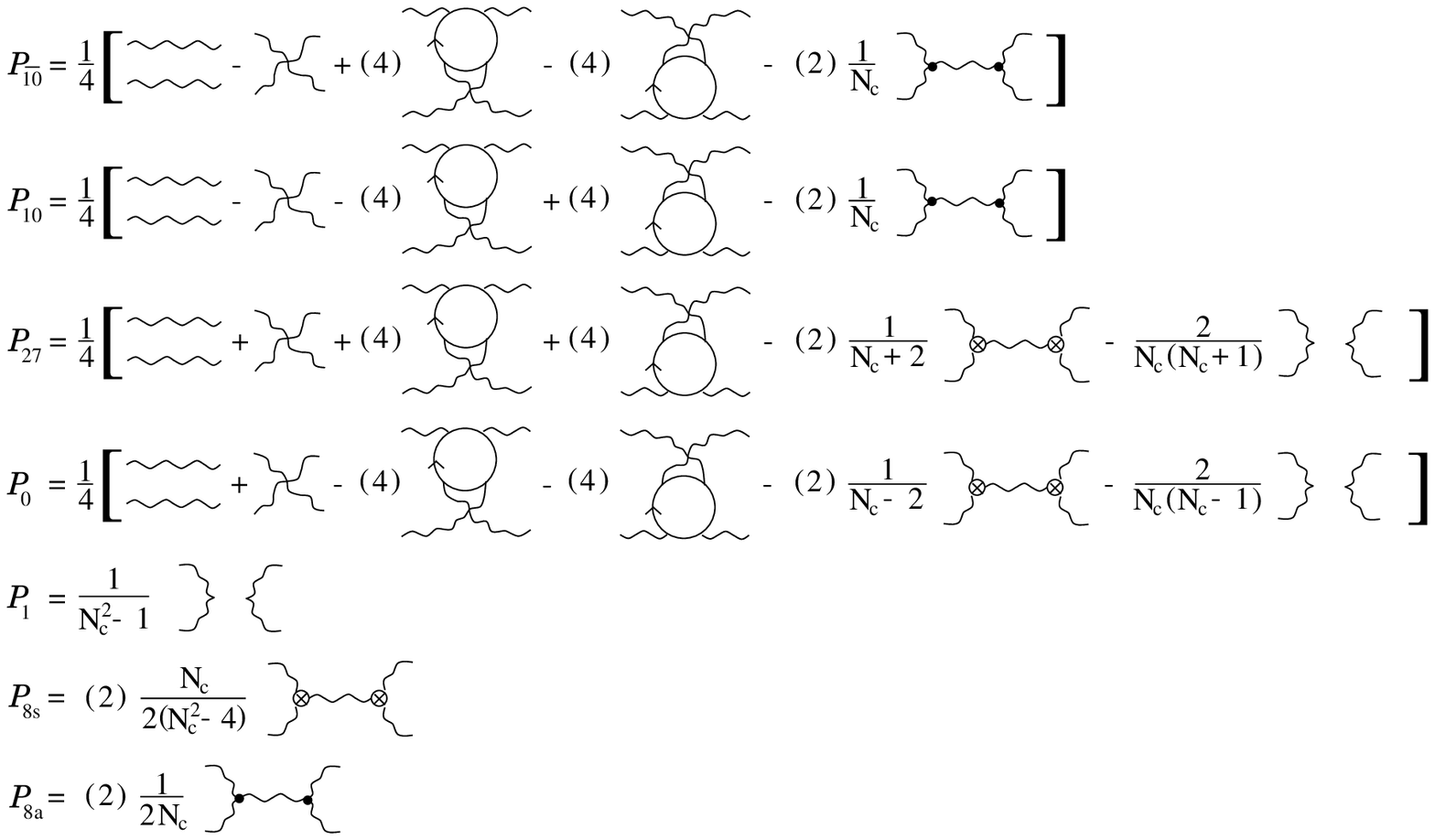}}
  }
  \caption{\em Projection operators needed to obtain the irreducible representations of the \mbox{\hspace{3em}} \mbox{\hspace{4em}} $8\otimes8=0\oplus1\oplus8_a\oplus8_s\oplus10\oplus\overline{10}\oplus27$ space. \mbox{\hspace{14em}} \mbox{\hspace{4em}}}
  \label{f:8x8operators}
\end{figure}

\begin{figure}[t]
  \hbox{\vbox{
    \begin{center}
    \mbox{\psfig{figure=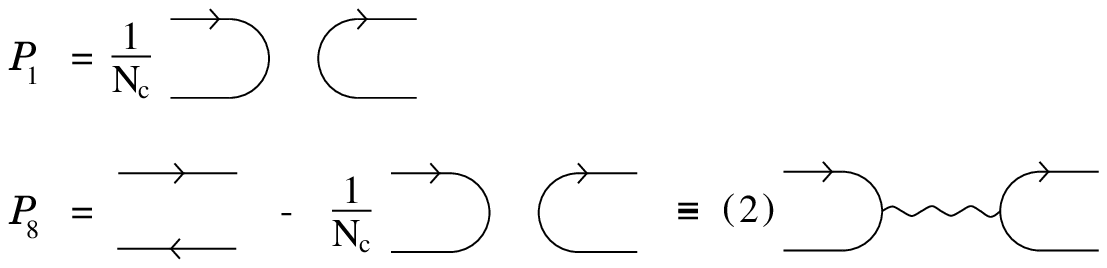}}
    \end{center}
  }}
  \caption{\em Projection operators needed to obtain the irreducible representations of the \mbox{\hspace{3em}} \mbox{\hspace{4em}} $3 \otimes \bar{3} = 1 \oplus 8$ space.  \mbox{\hspace{26em}} \mbox{\hspace{4em}}}
  \label{f:3x3_operators}
\end{figure}

\begin{figure}[t]
  \hbox{\vbox{
    \begin{center}
    \mbox{\psfig{figure=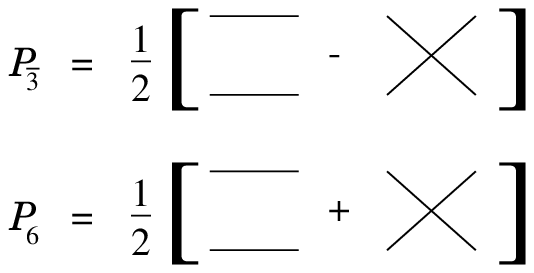}}
    \end{center}
  }}
  \caption{\em Projection operators needed to obtain the irreducible representations of the \mbox{\hspace{3em}} \mbox{\hspace{4em}} $3 \otimes 3 = \bar{3} \oplus 6$ space.}
  \label{f:3x3operators}
\end{figure}

\begin{figure}[t]
  \hbox{\vbox{
    \begin{center}
    \mbox{\psfig{figure=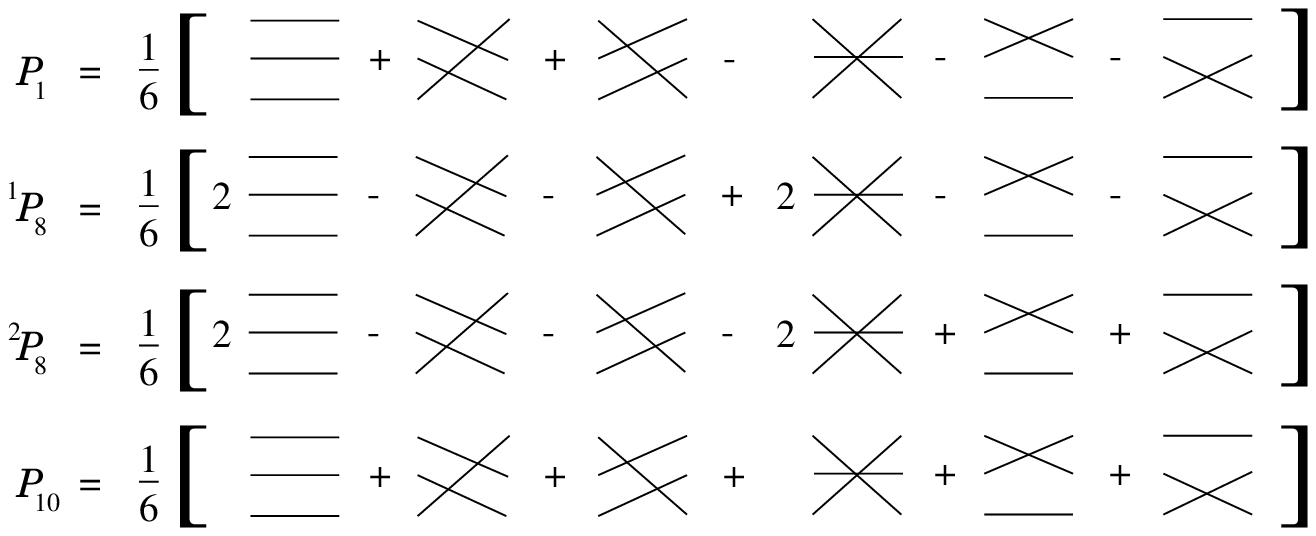}}
    \end{center}
  }}
  \caption{\em Projection operators needed to obtain the irreducible representations of the \mbox{\hspace{3em}} \mbox{\hspace{4em}} $3 \otimes 3 \otimes 3 = 1 \oplus 8 \oplus 8 \oplus 10$ space.}
  \label{f:3x3x3operators}
\end{figure}

In Figs~\ref{f:8x8operators}--\ref{f:3x3x3operators} we present the set of projection operators needed to obtain irreducible representations of the $8\otimes8$, $3\otimes\bar{3}$, $3\otimes3$, and $3\otimes 3\otimes3$ spaces, respectively. For the derivation of these operators we refer to Cvitanovi\'c~\cite{pc84}.

There are different normalizations adopted for gluons and the coupling constant $\alpha_s$. A difference in the overall normalization of the projection operators is not a problem as long as it is used consistently. The overall normalization is factorized out in most cases, as in the Examples~\ref{ss:ex2} and~\ref{ss:ex3}. In Example~\ref{ss:ex1} we have adopted the Gell-Mann (conventional) normalization of the gluons.

The difference in normalization, lies in how we interpret the curly lines, representing the gluons, and also in the interpretation of the triple vertices. With the Gell-Mann normalization, the gluon is interpreted as the octet projection in the $3\otimes \bar{3}$ space together with a factor 1/2 (compare Eq C11 of Appendix~\ref{a:rules}). This factor originates from the conventional QCD normalization of the generators of the SU(3) group, ie. 
\eqbe
  t^a = \frac{1}{2} \lambda^a.
\label{e:lambda}
\eqen

A more convenient way to normalize the gluon is to let the factor of $\frac{1}{2}$ in Eq~\ref{e:lambda} go into $\alpha_s$, ie. to let the gluon represent just the octet projection, without any factors. In the same spirit, the triple gluon vertices (symmetric and anti-symmetric) can be defined without any factors.

The program output comes with both normalizations. The answer which is expressed as a polynomial in $N_c$, is normalized with the second convention. If you use the polynomial ignoring the overall factor (and the numerical answer), you should replace the factors within parenthesis, in the operators in Figures~\ref{f:8x8operators} and~\ref{f:3x3_operators}, by a factor 1. If you prefer to use the Gell-Mann normalization, you must make use of the factors within parenthesis.

The projection operator, $P_0$, has the multiplicity 0 when the number of colours is 3. This subspace exists eg. in SU(2) where it has multiplicity 3 and corresponds to a real state.


\end{document}